\documentclass[
    aps,
    pra,
    longbibliography,
    superscriptaddress,amsmath,amssymb,amsfonts, 
    tightenlines,
    twocolumn,
    ]{revtex4-1}
    
\usepackage[utf8]{inputenc}  
\usepackage[T1]{fontenc}     
\usepackage[english]{babel}  
\usepackage{hyperref}  
\hypersetup{
    colorlinks=true,
    allcolors=purple,
}
\usepackage{graphicx} 
\usepackage[babel]{microtype}  
\usepackage{amsmath,amssymb,amsthm,bm,amsfonts,mathrsfs,bbm} 
\usepackage{setspace}
\usepackage{times}
\usepackage{braket}
\usepackage[bottom]{footmisc}
\usepackage{blochsphere}
\usepackage{pgfplots}
\usepackage[normalem]{ulem}
\usepackage{lipsum}
\usepackage{natbib}
\usepackage{bm}
\usepackage{eucal}
\usepackage{pifont}
\usepackage{xcolor}
\usepackage{tikz}
\usepackage{setspace}
\usetikzlibrary{quantikz2}
\usepackage[skins,breakable]{tcolorbox}
\tcbuselibrary{listings}
\usepackage{tcolorbox}
 
\pgfplotsset{compat=1.17}
\usetikzlibrary{calc,3d,shapes, pgfplots.external, intersections}
\usepackage{float}
\makeatletter
\let\newfloat\newfloat@ltx
\makeatother
\usepackage{algorithm}
\usepackage{listings}
\usepackage{physics}
\usepackage[noend]{algorithmic}
\usepackage{eqparbox}

\usepackage{physics}
\usepackage{tikz}
\usepackage{mathtools}
\usetikzlibrary{arrows.meta}
\usepackage{csquotes}

\usepackage{xspace}  
\usepackage{pgfplots}
\usepackage{verbatim}
\usepackage{amsmath}

\newcommand\id{\leavevmode\hbox{\small1\kern-3.3pt\normalsize1}}

\usepackage{hyperref}
\usepackage{booktabs}
\usepackage{multirow}    
\usepackage{array}

\usepackage{xcolor}
\usepackage{pdfpages}
\usepackage{pgffor}
\makeatletter
\AtBeginDocument{\let\LS@rot\@undefined}
\makeatother

\definecolor{backcolour}{rgb}{0.95,0.95,0.92}

\lstdefinestyle{mystyle}{
    backgroundcolor=\color{backcolour},   
    basicstyle=\ttfamily\footnotesize,
    breakatwhitespace=false,         
    breaklines=true,                 
    captionpos=b,                    
    keepspaces=true,                 
    numbers=none,                    
    numbersep=5pt,                  
    showspaces=false,                
    showstringspaces=false,
    showtabs=false,                  
    tabsize=2
}

\begin{document}

\title{\texttt{Q-Newton}: Hybrid Quantum-Classical Scheduling for\\Accelerating Neural Network Training with Newton's Gradient Descent}

\newcommand*{\affaddr}[1]{\textit{#1}}
\newcommand*{\affmark}[1][*]{\textsuperscript{#1}}

\author{
  Pingzhi Li\affmark[1], Junyu Liu\affmark[2,3,4], Hanrui Wang\affmark[5], and Tianlong Chen\affmark[1]\\
  \affaddr{\affmark[1]Department of Computer Science, The University of North Carolina at Chapel Hill, Chapel Hill, NC 27599, USA}\\
  \affaddr{\affmark[2]Department of Computer Science, The University of Pittsburgh, Pittsburgh, PA 15260, USA}\\
  \affaddr{\affmark[3]Pritzker School of Molecular Engineering, The University of Chicago, Chicago, IL 60637, USA}\\
  \affaddr{\affmark[4]Department of Computer Science, The University of Chicago, Chicago, IL 60637, USA}\\
  \affaddr{\affmark[5]Computer Science Department, University of California Los Angeles, Los Angeles, CA 90095, USA}\\
}





\maketitle

{\bf {Training deep neural networks with second-order optimization methods offers faster convergence but remains computationally prohibitive due to the cubic-time complexity of matrix inversion. Recent advancements in quantum computing promise exponential acceleration for specific linear algebra tasks, yet their practical integration into machine learning pipelines faces significant challenges. Here, we introduce \texttt{Q-Newton}, a hybrid quantum-classical optimization framework that dynamically schedules matrix inversion operations between quantum and classical processors based on real-time matrix properties. By combining Hessian regularization, sparsity-aware pruning, and condition number estimation, \texttt{Q-Newton} reduces training time by up to $90\%$ compared to classical second-order methods without compromising model accuracy. Our approach demonstrates that selectively offloading well-conditioned, sparse matrices to quantum linear solvers while processing ill-conditioned (\textit{e.g.} condition number $\kappa>10^4$) matrices classically maximizes computational efficiency. We show that Hessians naturally evolve toward quantum-favorable properties during training (\textit{e.g.} $\kappa$ drops to $1/1000\times$), enabling increasingly effective quantum acceleration. These findings prompt a practical pathway for leveraging quantum computing's advantages within existing deep learning frameworks and provide insights into quantum-classical co-design principles for machine learning model training. Our code is available at \href{https://github.com/UNITES-Lab/q-newton}{github.com/UNITES-Lab/q-newton}.}}

\section{Introduction}

Training deep neural networks~(\textit{e.g.} large language models~(LLMs)) remains computationally demanding despite significant advancements in hardware accelerators and optimization techniques~\citep{kingma2014adam, choquette2020nvidia,peccerillo2022survey}. First-order methods such as stochastic gradient descent~(SGD)~\citep{robbins1951stochastic} and its variants~\citep{kingma2014adam,loshchilov2019decoupledweightdecayregularization,zhao2024galore} dominate modern deep learning due to their computational efficiency and scalability~\cite{sun2020optimization}. However, these approaches often converge slowly on large neural networks (\textit{e.g.} large language models), suffer from sensitivity to hyperparameter tuning, and frequently struggle with challenging loss landscapes~\cite{reddi2019convergence}. This computational bottleneck has become increasingly evident as model sizes continue to grow exponentially, with state-of-the-art models now containing hundreds of billions of parameters~\cite{liu2024deepseek,achiam2023gpt,grattafiori2024llama}.

Second-order optimization methods, particularly Newton's gradient descent~(Newton's GD) and its variants~\citep{more1982newton,polyak2007newton}, offer substantially faster convergence by incorporating curvature information from the Hessian matrix~\cite{nocedal2006large}. By accounting for the relationships between parameters, these approaches can achieve superlinear or even quadratic convergence rates compared to the linear rates of first-order methods~\cite{roosta2019sub}. However, their practical adoption remains limited due to the $\mathcal{O}(n^3)$ time complexity required for inverting the $n \times n$ Hessian matrix at each iteration~\cite{martens2010deep}. This cubic scaling renders traditional second-order methods impractical for modern neural networks, especially large language models~(LLMs), with billions or trillions of parameters, despite their theoretical advantages.

\begin{figure}
    \centering
    \includegraphics[width=0.95\linewidth]{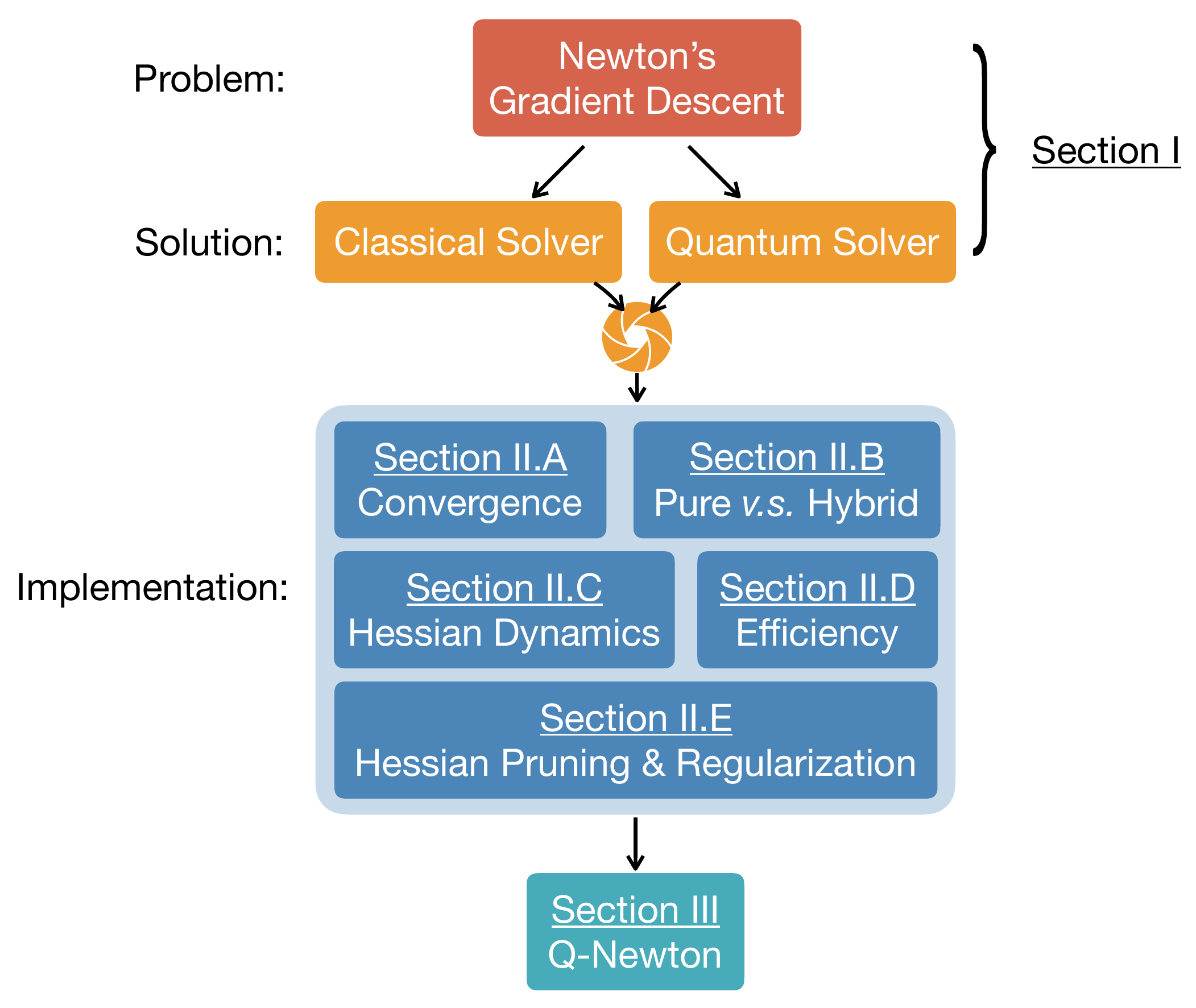}    
    \vspace{-10pt}
    \caption{\textbf{Overview.} We propose the hybrid quantum-classical solver of Newton's gradient descent as \texttt{Q-Newton}, a general neural network training framework. Starting from Newton's gradient descent problem, we develop solutions via both classical and quantum solvers, integrating them through our hybrid approach. We evaluate \texttt{Q-Newton} along five dimensions: convergence improvements over first-order methods, comparative analysis between pure quantum/classical and hybrid implementations, Hessian matrix dynamics throughout training, computational efficiency across different neural architectures, and the effects of Hessian pruning and regularization techniques. Through these investigations, \texttt{Q-Newton} demonstrates significant training acceleration capabilities by dynamically leveraging the strengths of quantum and classical computation for matrix inversion operations.}
    \label{fig:main-results}
\end{figure}

Fortunately, quantum computing offers a potential solution for this challenge through quantum linear system algorithms~(QLSAs)~\cite{harrow2009quantum, childs2018toward}. QLSAs promise exponential speedups for matrix inversion operations under specific conditions, reducing complexity from $\mathcal{O}(n^3)$ to $\mathcal{O}(\text{polylog}(n))$ for well-conditioned, sparse matrices~\cite{harrow2009quantum}. However, this theoretical advantage comes with important caveats: quantum speedups are highly sensitive to the condition number ($\kappa$) and sparsity ($d$) of the target matrix, with actual runtime scaling as $\mathcal{O}(d \cdot \kappa \cdot \text{polylog}(n))$~\cite{childs2018toward, dervovic2018quantum}. For ill-conditioned or dense matrices commonly encountered in modern neural network training~\citep{saarinen1993ill,yao2020pyhessian}, these quantum algorithms may provide little or no advantage over classical approaches.

In this work, we introduce \texttt{Q-Newton}, a hybrid quantum-classical scheduler for Newton's method that dynamically determines the optimal processor (\textit{i.e.}~quantum or classical) for each matrix inversion operation during neural network training. Rather than treating quantum computing as an unconditional accelerator, \texttt{Q-Newton} implements an intelligent scheduling policy based on real-time estimation of matrix properties, along with techniques to actively reshape the Hessian matrix to favor QLSAs. Our system combines three key innovations: ($1$)~an efficient condition number and sparsity estimator for large matrices, ($2$)~an adaptive Hessian regularization and pruning framework that preserves critical curvature information while improving quantum compatibility, and ($3$)~a cost-aware scheduler that dynamically routes inversion operations to maximize computational efficiency.

Through extensive experiments on three popular neural network architectures (DNN, BERT, GPT) across two major machine learning tasks (computer vision and natural language processing), we demonstrate that \texttt{Q-Newton} reduces training time by up to $90\%$ compared to classical second-order methods without compromising model accuracy. Our results reveal that neural network Hessians naturally evolve toward more favorable properties for QLSAs as training progresses, with condition numbers decreasing and structural sparsity emerging over time. This observation leads to a natural scheduling pattern where classical processors handle early training iterations with ill-conditioned Hessians, gradually transitioning to quantum processing as matrices become more amenable to QLSAs. Beyond performance improvements, \texttt{Q-Newton} provides valuable insights into the practical boundaries between quantum \& classical computing for accelerating machine learning. Our framework offers a promising path towards practical quantum-classical integration that maximizes the strengths of both paradigms while minimizing their respective limitations.

\section{Results}

\begin{table}[htbp]
  \centering
  \caption{Summary of training runtime evaluation results across two pairs of models and datasets. Newton's gradient descent methods exhibit significantly better convergence, requiring fewer training steps to achieve the same performance compared to SGD. \texttt{Q-Newton} demonstrates significant superiority in training time.}
   \resizebox{0.99\linewidth}{!}{
    \begin{tabular}{l|rrr|rrr}
    \toprule
    \midrule
    \multirow{2}{*}{Method} & \multicolumn{3}{c|}{DNN (MNIST)} & \multicolumn{3}{c}{GPT (WikiText)} \\
    \cmidrule{2-7}
     &  Steps & Time/Step &  Time &  Steps & Time/Step &  Time \\
    \midrule
    SGD & $1500$ & $0.1$s & $204$s & $6200$ & $0.1$s & $11.1$min \\
    \midrule
    Classical-Only &  $1000$ & $1.1$s & $1064$s & $2000$ & $9.8$s & $334.2$min \\
    Quantum-Only &  $1000$ & $0.2$s & $170$s & $2000$ & $0.1$s & $3.8$min \\
    \texttt{Q-Newton} &  $1000$ & $0.1$s & $123$s & $2000$ & $0.1$s & $3.5$min \\
    \midrule
    \bottomrule
    \end{tabular}}
  \label{tab:combined-results}
\end{table}



\begin{figure*}
    \centering
    \includegraphics[width=0.98\textwidth]{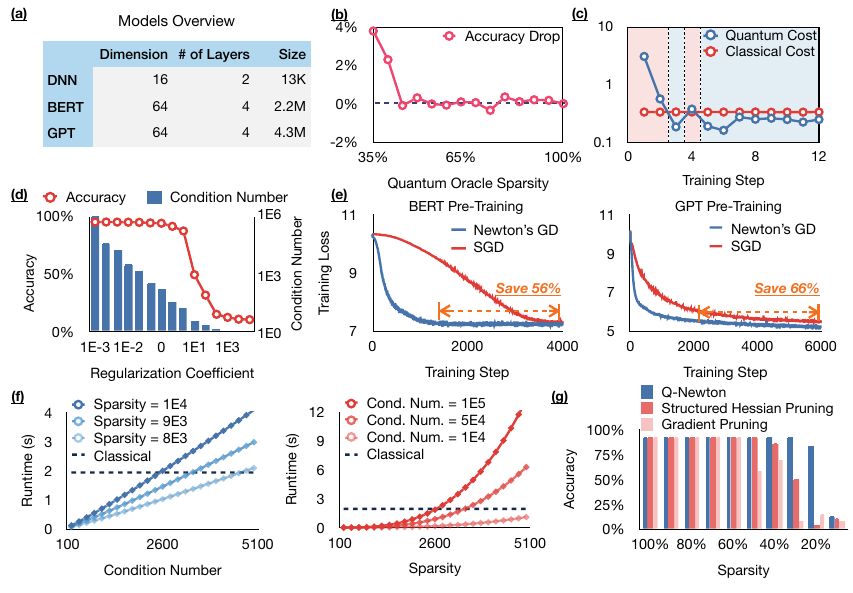}    
    \vspace{-10pt}
    \caption{\textbf{Performance evaluation and ablation study of \texttt{Q-Newton}.} (\textbf{a}) Overview of neural network models used for evaluation, showing architecture dimensions, layer counts, and parameter sizes. (\textbf{b}) Impact of quantum oracle sparsity on model accuracy, demonstrating minimal accuracy drop ($<0.5\%)$ with up to $65\%$ sparsification. (\textbf{c}) Comparison of computational costs between quantum and classical solvers across training steps, showing the dynamic scheduling decisions of \texttt{Q-Newton} with quantum advantage emerging after step $4$. (\textbf{d}) Relationship between regularization coefficient, model accuracy (red line), and Hessian condition number (blue bars), highlighting the optimal regularization range for maintaining accuracy while improving matrix conditioning. (\textbf{e}) Convergence comparison between Newton's gradient descent and SGD on BERT and GPT pre-training tasks, demonstrating $56\%$ and $66\%$ training step reduction, respectively. (\textbf{f}) Runtime analysis of quantum solver with varying sparsity levels against classical solver (dashed line) as condition number increases, showing crossover points where quantum advantage emerges. (\textbf{g}) Comparative accuracy evaluation of \texttt{Q-Newton} against alternative pruning strategies across different sparsity levels, showing \texttt{Q-Newton}'s superior preservation of model performance at high sparsity levels.}
    \label{fig:main-results}
\end{figure*}

\subsection{Second-order methods improve training convergence}

Second-order optimization methods leverage curvature information to accelerate neural network training significantly beyond first-order approaches. Table~\ref{tab:combined-results} and Figure~\ref{fig:main-results}~\textbf{\uline{(e)}} show the superior convergence properties of Newton's GD compared to SGD across three network architectures. When applied to finetuning a BERT~\cite{devlin-etal-2019-bert} model, Newton's GD achieves convergence in only $44\%$~(\textit{i.e.}, saving $56\%$) of the iterations required by SGD. Similarly, for the GPT model, Newton's GD reduces the number of training steps by up to $66\%$ while achieving better final performance.

However, these convergence improvements come at a substantial computational cost. Our measurements in Table~\ref{tab:combined-results} indicate that classical implementations of Newton's GD require up to $98\times$ longer per-iteration computation time compared to SGD when applied to modern neural networks with millions of parameters. This prohibitive overhead explains why, despite their theoretical advantages, second-order methods remain underutilized in modern deep learning practice~\citep{roosta2019sub,goldfarb2021practicalquasinewtonmethodstraining}. The computational bottleneck lies specifically in the $\mathcal{O}(n^3)$ complexity of inverting the Hessian matrix, which dominates the overall training time as model size increases.

\subsection{Pure quantum or classical approaches are suboptimal for matrix inversion}
We conducted a comprehensive analysis comparing classical and quantum matrix inversion across several matrix properties to identify their respective domains of advantage. Figure~\ref{fig:main-results}~\textbf{\uline{(c)}} demonstrates that neither approach consistently outperforms the other across all neural network training steps.

Specifically, for classical inversion using LU decomposition~\citep{anderson1999lapack,harris2020array}, execution time scales as $\mathcal{O}(n^3)$ regardless of the matrix's other properties. In contrast, QLSAs exhibit non-trivial performance characteristics that depend critically on matrix sparsity and condition number. Our empirical measurements, shown in Figure~\ref{fig:main-results}~\textbf{\uline{(f)}}, confirm that QLSA execution time scales approximately as $\mathcal{O}(d \cdot \kappa \cdot \log(n))$, where $d$ represents the density factor (proportion of non-zero elements per row), $\kappa$ is the condition number, and $n$ is the matrix dimension.

Therefore, quantum approaches demonstrate clear advantages for \textit{large}, \textit{well-conditioned}, and \textit{sparse} matrices. Conversely, this advantage rapidly diminishes for ill-conditioned or dense matrices. Figure~\ref{fig:main-results}~\textbf{\uline{(f)}} identifies the "crossover boundaries" where quantum and classical approaches achieve equivalent performance. For example, matrices with the condition number above $2600$ and density greater than $30\%$ are typically processed more efficiently using classical methods.

In summary, these findings indicate that a binary choice between quantum or classical computation is inherently suboptimal for neural network training, where matrix properties (\textit{i.e.} sparsity and condition number) vary substantially across training iterations and network layers. Instead, a hybrid approach that dynamically selects the appropriate processor based on certain matrix characteristics presents a more effective way.

\subsection{Neural network Hessians evolve favorably during training}

\begin{figure}[t]
    \centering
    \includegraphics[width=0.8\linewidth]{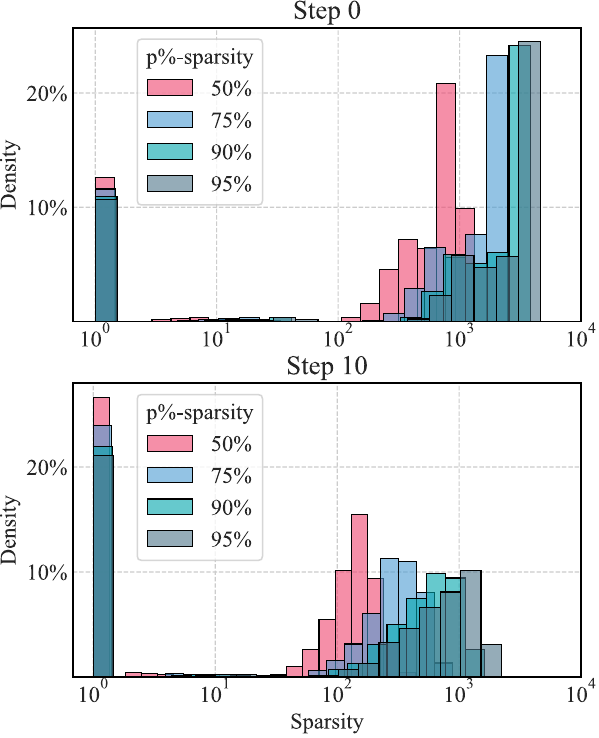}
    \caption{Distribution of $p\%$ quantum sparsity for Hessian matrix at training steps $0$ (up) and $10$ (down). $p\%$-sparsity represents the minimum number of elements in a row summing to $p\%$ of its total absolute magnitude. Data from DNN training on MNIST (Hessian size: $12544\times 12544$), showing Hessian magnitude distribution evolution. We evaluate the $p\%$ across $\{50\%, 75\%, 90\%, 95\%\}$.}
    \label{fig:hessian-distribution}
\end{figure}

One key observation enabling effective hybrid scheduling is that the properties of neural network Hessian matrices naturally evolve throughout the training process. Figure~\ref{fig:hessian-distribution} shows the sparsity patterns of Hessians extracted during DNN training at step $0$ \textit{v.s.} $10$, illustrating the sparsifying trend along training.

Our empirical results show that, in early training phases, Hessian matrices typically exhibit high condition numbers (\textit{e.g.} $\kappa > 10^4$) and relatively dense structures, making them poor candidates for quantum acceleration. This aligns with previous findings that loss landscapes are particularly steep and irregular during initial training steps~\cite{keskar2016large}. As training progresses, we observe that Hessian matrices gradually develop more favorable properties for QLSAs: condition numbers drop to $1/10\times$-$1/1000\times$ while natural sparsity patterns emerge~\citep{yao2020pyhessian}.



\subsection{Dynamic hybrid scheduling maximizes computational efficiency}

Building on these observations, \texttt{Q-Newton} implements a dynamic scheduling policy that routes matrix inversion operations to either quantum or classical processors based on real-time matrix properties. Figure~\ref{fig:main-results}~\textbf{\uline{(c)}} shows \texttt{Q-Newton} scheduler's decisions throughout the training steps of DNN training, demonstrating that classical processing dominates early training phases, with quantum processing gradually assuming a larger role as matrices become more QLSA-friendly.

Our cost scheduler balances the tradeoffs between classical computation scaling as $\mathcal{O}(n^3)$ and quantum computation scaling as $\mathcal{O}(d \cdot \kappa \cdot \log(n))$. By incorporating light-weight real-time estimates of condition number and sparsity in \texttt{Q-Newton}, the scheduler can consistently identify the most efficient processor for each inversion operation, avoiding the pitfalls of static allocation to either computing paradigm.

Table~\ref{tab:combined-results} demonstrates the performance advantages of this hybrid approach compared to pure classical or pure quantum implementations. Across all tested network architectures, \texttt{Q-Newton} achieves $8$-$100\times$ speedup over classical Newton's GD and $1.1$-$1.4\times$ speedup over pure quantum approaches. This performance gap potentially widens for larger models, where the matrix properties exhibit greater variability and the potential for targeted acceleration increases.

\subsection{Hessian pruning and regularization enhance quantum compatibility}

While neural network Hessians naturally evolve toward quantum-favorable properties, \texttt{Q-Newton} further improves quantum compatibility through targeted pruning and regularization techniques. Figure~\ref{fig:main-results}~\textbf{\uline{(b)}} \& \textbf{\uline{(d)}} demonstrates that carefully designed sparsification strategies can substantially improve computational efficiency without degrading optimization performance (\textit{e.g.} sparsity $>40\%$, regularization coefficient $<1$).

Specifically, our magnitude-based, symmetry-aware pruning approach removes small Hessian elements while preserving the matrix's positive-definiteness and critical eigenstructure. By applying pruning thresholds calibrated to each matrix's magnitude distribution, we achieve up to $70\%$ sparsity with negligible impact on convergence rate or final model accuracy, as shown in Figure~\ref{fig:main-results}~\textbf{\uline{(b)}}. Similarly, Figure~\ref{fig:main-results}~\textbf{\uline{(d)}} shows that our regularization approach reduces condition numbers by up to $75\%$ while preserving the essential curvature information that drives optimization and maintains accuracy.

\section{Methods}

\begin{figure*}
    \centering
    \includegraphics[width=0.85\textwidth]{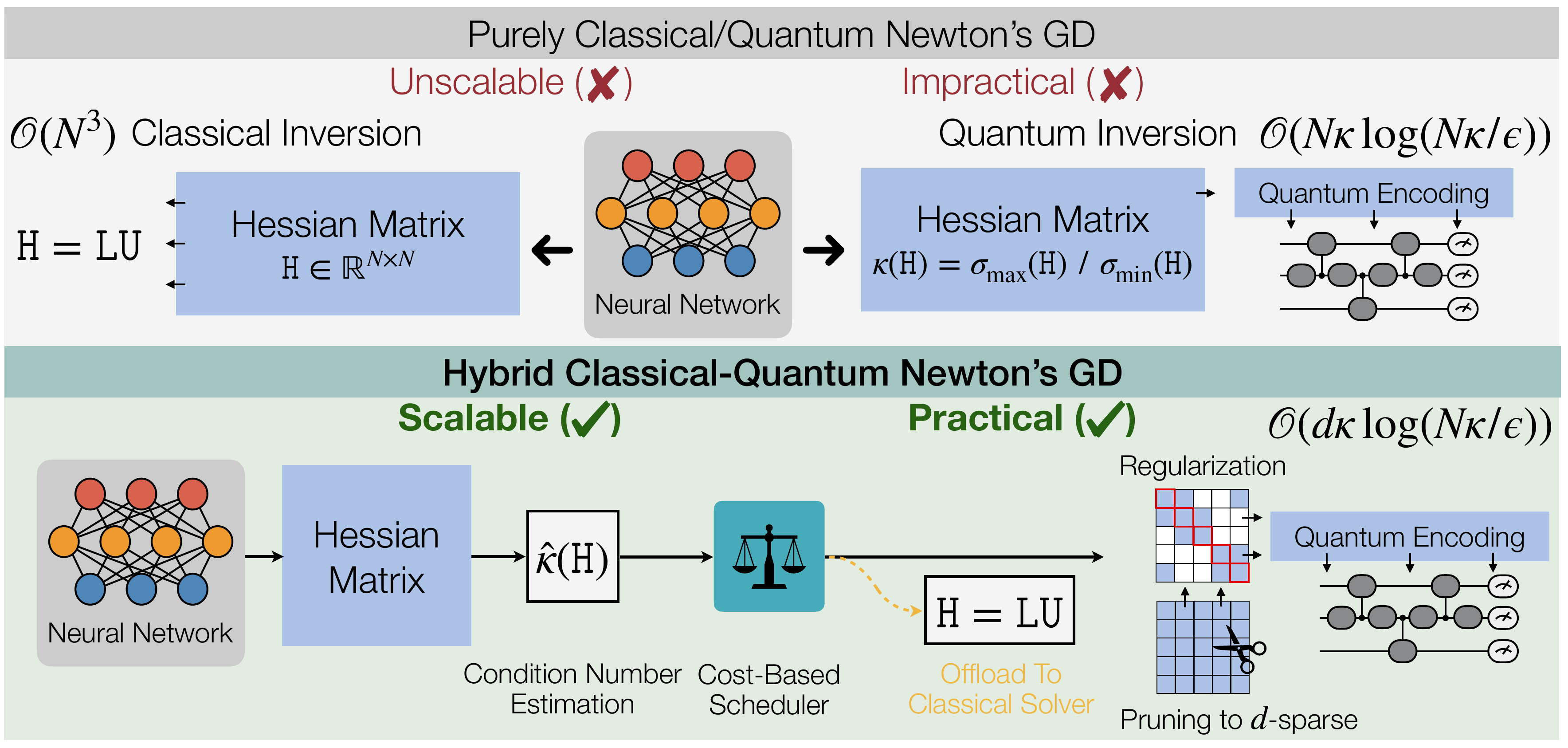}    
    \vspace{-10pt}
    \caption{\textbf{Overview of \texttt{Q-Newton}}. Both \uline{purely classical and quantum Newton's GD} are unscalable and impractical. Purely classical Newton's GD calculates the Hessian inversion via LU decomposition \textit{etc.}, with a time complexity of $\mathcal{O}(N^3)$. Quantum Newton's GD requires substantial time when the Hessian is either ill-conditioned or non-sparse. \uline{Hybrid classical-quantum Newton's GD} adaptively schedules between classical and quantum inversion solvers, and incorporates pruning and regularization techniques, making it scalable and practical.}
    \label{fig:pipeline}
\end{figure*}

\subsection{Hessian estimation and approximation}

\begin{figure}[htb]
    \centering
    \includegraphics[width=1.02\linewidth]{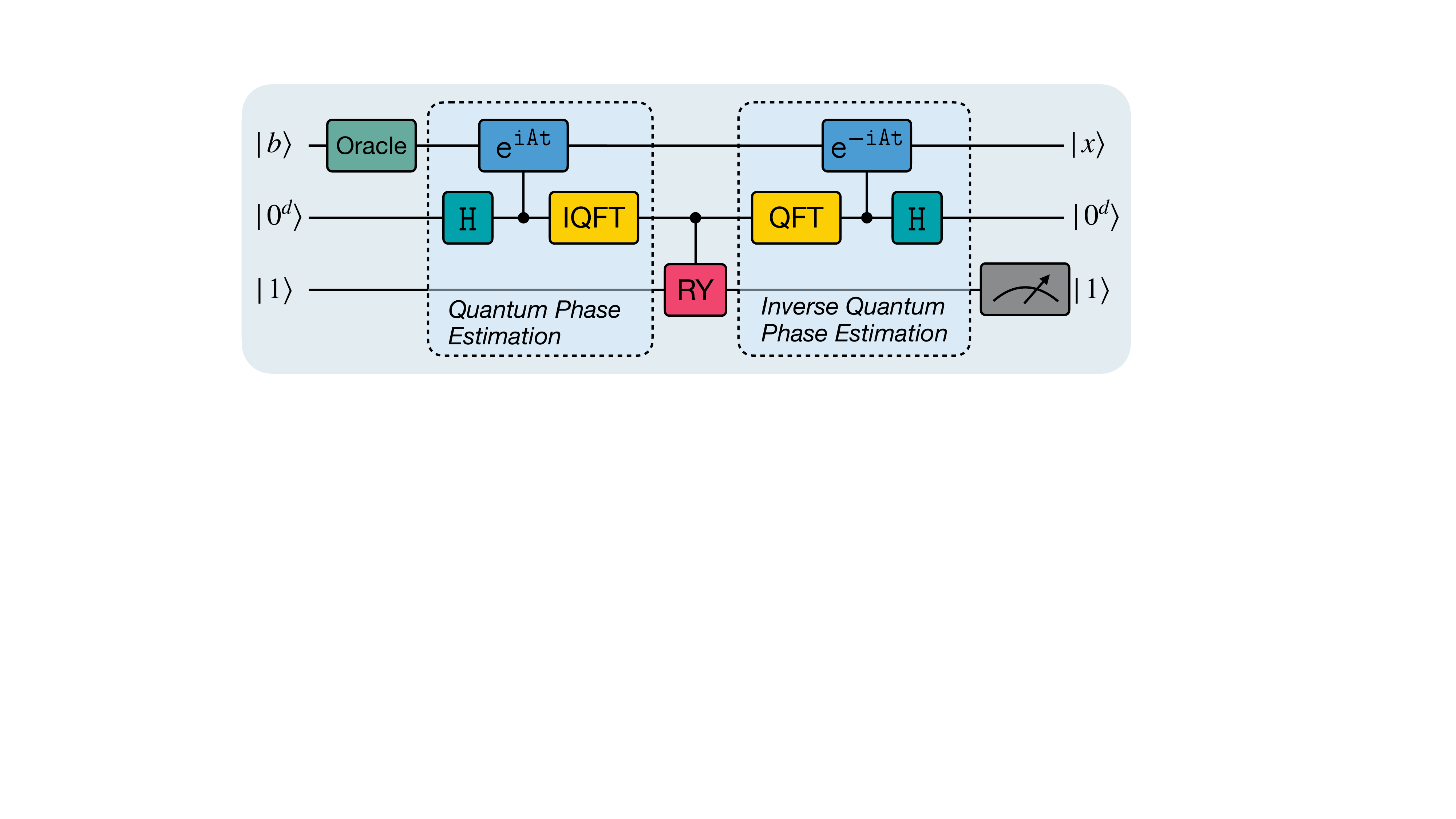}
    \caption{Schematic of the QLSA circuit flowing from left to right. }
    \label{fig:hhl-circuit}
\end{figure}

Computing the full Hessian matrix for large neural networks is prohibitively expensive, requiring $\mathcal{O}(n^2)$ backpropagation passes, where $n$ is the number of parameters. To make second-order optimization practical, we employ a block-diagonal Hessian approximation that captures parameter interactions within each network layer while ignoring cross-layer dependencies~\citep{martens2015optimizing}.

Specifically, for each network layer $l$ with parameter count $n_l$, we compute the layer-wise Hessian $\mathbf{H}_l \in \mathbb{R}^{n_l \times n_l}$ using efficient vector-Hessian product operations~\cite{pearlmutter1994fast}:
\begin{align}
\mathbf{H}_l = \frac{1}{|B|} \sum_{i \in B} \nabla^2_{\theta_l} \mathcal{L}(x_i, y_i; \theta),
\end{align}
where $B$ represents a mini-batch of training examples, $\mathcal{L}$ is the loss function, and $\theta_l$ are the parameters of layer $l$. This block-diagonal structure enables parallel and efficient computing of layer-wise Hessians and aligns naturally with the layerwise structure of most neural networks.

To further reduce computational overhead, we implement an efficient condition number estimator that avoids explicitly computing the full eigendecomposition. Instead, we use the Merikoski bound~\citep{merikoski1997best} to estimate the condition number without computing eigenvalues directly:
\begin{align}
\kappa_2(A) \leq \frac{1+x}{1-x}, \quad x = \sqrt{1-(n/\text{trace}(A))^n \det(A)}
\end{align}

\subsection{Hessian pruning and regularization}

Our Hessian pruning strategy builds on magnitude-based sparsification while incorporating additional constraints to preserve symmetry properties. 
To preserve symmetry, we apply pruning simultaneously to corresponding elements $\mathbf{H}_{ij}$ and $\mathbf{H}_{ji}$. Additionally, we enforce a structural constraint that maintains the positive-definiteness of the pruned Hessian by ensuring that all eigenvalues remain positive. When this constraint would be violated by standard magnitude pruning, we adjust the sparsification pattern to preferentially retain elements that contribute to the matrix's positive-definiteness.

Complementing pruning, our adaptive regularization approach further modifies the eigenvalue distribution to improve conditioning while preserving optimization performance. Specifically, given a regularization coefficient $\epsilon$, we compute the regularized Hessian matrix $\mathbf{H}_{\text{reg}} \in \mathbb{R}^{n\times n}$ as follows:
\begin{align*}
\mathbf{H}_\text{reg} = \mathbf{H} + \epsilon * \mathbf{I},
\end{align*}
where $\mathbf{I} \in \mathbb{R}^{n\times n}$ is the identity matrix.

\subsection{Cost estimation and scheduling policy}

The core of \texttt{Q-Newton}'s scheduling mechanism is to estimate execution time for both classical and quantum matrix inversion. For classical inversion using LU decomposition, the cost scales as:
\begin{align}
T_{\text{classical}}(n) = c_1 \cdot n^3 + c_2,
\end{align}
where $c_1$ and $c_2$ are hardware-specific constants determined through empirical calibration.

For quantum linear system algorithms, we model the cost:
\begin{align}
T_{\text{quantum}}(n, \kappa, d) = q_1 \cdot d \cdot \kappa \cdot \log(n/\epsilon) + q_2,
\end{align}
where $d$ is the density factor (proportion of non-zero elements), $\kappa$ is the condition number, $\epsilon$ is the desired solution precision, and $q_1$ and $q_2$ are quantum hardware-specific constants, where we assume quantum gate speed is possibly improved by
Attosecond Physics~\citep{zhong2020quantum}, towards a possible gate time
of attoseconds (\textit{i.e.}, ~$10^{-18}$s).

At each optimization step, the scheduler computes these cost estimates for each layer's Hessian matrix and routes the inversion operation to the processor with the lower predicted execution time. This decision takes into account the current properties of each matrix, including dimension $n$, estimated condition number $\kappa'$, and sparsity $d'$ pattern after pruning. The scheduling algorithm can be summarized as:
\begin{align}
\text{Processor} = 
\begin{cases}
\text{Quantum}, & \text{if}\ T_{\text{quantum}}(n, \kappa, d) < T_{\text{classical}}(n) \\
\text{Classical}, & \text{otherwise}
\end{cases}
\end{align}

\subsection{Implementation and experimental setup}

We implemented and simulated \texttt{Q-Newton} using PyTorch for neural network operations. For QLSAs, we employed the block-encoding approach~\citep{jennings2023efficient} for simulation, which offers improved scaling with respect to the condition number compared to the original HHL algorithm~\citep{harrow2009quantum}. 

To evaluate \texttt{Q-Newton}'s performance across various neural network architectures, we trained multiple model types, including DNN on MNIST, BERT \& GPT models on the WikiText dataset. The detailed hyper-parameters are shown in Figure~\ref{fig:main-results}~\textbf{\uline{(a)}}. For each architecture, we compared \texttt{Q-Newton} against several baselines: ($1$) standard SGD, ($2$) pure classical  Newton's GD, and ($3$) pure quantum Newton's GD. All models were trained to convergence, with hyperparameters individually tuned for each optimizer to ensure fair comparison.

\section{Discussion}

\subsection{Scientific implications}

Our results demonstrate that hybrid quantum-classical computing can make second-order optimization practical for neural network training, attempting to address a long-standing computational bottleneck in second-order machine learning~\citep{pearlmutter1994fast,sun2020optimization}. By selectively applying quantum acceleration to well-conditioned, sparse matrices while processing ill-conditioned matrices classically, \texttt{Q-Newton} achieves the convergence benefits of second-order methods without their prohibitive computational costs. This approach exemplifies how quantum computing can enhance, rather than replace, classical algorithms in domains where neither paradigm is universally superior.

Beyond the specific application to Newton's GD, our work provides insights into the broader potential for quantum-classical algorithms in machine learning. The observation that neural network Hessians evolve toward quantum-amenable properties suggests that similar hybrid approaches could accelerate other matrix-intensive operations in deep learning, such as attention mechanisms in Transformers~\citep{vaswani2017attention}, singular value decomposition in recommendation systems~\citep{rendle2010factorization}, or kernel operations in Gaussian processes~\citep{seeger2004gaussian}.

The Hessian analysis techniques developed for \texttt{Q-Newton} also offer new tools for understanding neural network optimization dynamics. Our findings regarding the natural evolution of condition numbers and sparsity patterns during training contribute to the growing body of research on loss landscape geometry and its relationship to generalization performance. These insights could inform the development of new optimization algorithms and initialization strategies specifically designed to promote well-conditioned loss surfaces.

\subsection{Limitations and future directions}

Several limitations of our current approach suggest directions for future research. \uline{First}, our implementation relies on simulation rather than actual quantum hardware, abstracting away some practical challenges of NISQ devices. Adapting \texttt{Q-Newton} to account for realistic error rates, limited qubit connectivity, and decoherence effects represents an important next step toward practical deployment.

\uline{Second}, while our block-diagonal Hessian approximation makes second-order optimization tractable, it ignores potentially important cross-layer parameter interactions. More sophisticated Hessian approximation techniques, \textit{e.g.} Kronecker-factored approaches~\cite{martens2015optimizing}, could further improve optimization performance while maintaining computational feasibility.

\uline{Finally}, extending our hybrid approach to other second-order methods such as quasi-Newton approaches (\textit{e.g.}, L-BFGS) or natural gradient descent could offer additional performance improvements, particularly for problems where exact Hessian computation remains prohibitive.

\section{Conclusion and Outlook}

This work introduces \texttt{Q-Newton}, a hybrid quantum-classical framework that makes second-order optimization (\textit{i.e.} Newton's gradient descent) practical for training modern neural networks. By scheduling matrix inversion operations between quantum and classical processors based on real-time Hessian properties, our approach achieves up to $90\%$ reduction in training time compared to classical second-order methods while maintaining model accuracy. Our findings demonstrate that neural network Hessians naturally evolve toward quantum-amenable properties during training, enabling increasingly effective quantum acceleration as training progresses.

\texttt{Q-Newton} represents a significant departure from the conventional binary view of quantum versus classical computing. Rather than treating quantum computing as an unconditional accelerator, our results reveal that co-design principles can maximize the advantages of both paradigms while minimizing their respective limitations. 

Several challenges remain to be addressed in future work. Current implementations still rely on the simulation of QLSAs rather than deployment on actual quantum hardware, which could introduce additional constraints from noise and decoherence. Extending our approach to increasingly large neural networks, particularly those with billions or trillions of parameters, will require further innovations in Hessian approximation and quantum resource estimation. Additionally, exploring the application of hybrid scheduling to other matrix-intensive operations in deep learning, \textit{e.g.} attention mechanisms in transformers, represents a promising direction for broader impact.

Looking forward, \texttt{Q-Newton} establishes a practical pathway for integrating quantum computing capabilities into existing machine learning workflows. Rather than expecting fault-tolerant quantum computing to deliver universal advantages, our results show that carefully targeted quantum acceleration can potentially provide substantial benefits in the near term. 

\section*{Acknowledgment}
JL is supported in part by International Business Machines (IBM) Quantum through the Chicago Quantum Exchange, and the Pritzker School of Molecular Engineering at the University of Chicago through AFOSR MURI (FA9550-21-1-0209). JL is also supported in part by the University of Pittsburgh, School of Computing and Information, Department of Computer Science, Pitt Cyber, PQI Community Collaboration Awards, and by NASA under award number 80NSSC25M7057. This research used resources of the Oak Ridge Leadership Computing Facility, which is a DOE Office of Science User Facility supported under Contract DE-AC05-00OR22725. PL and TC are supported in part by Cisco Faculty Award and UNC SDSS Seed Grant.

\bibliographystyle{naturemag}

\section*{Appendix}

\subsection*{A. Quantum Superiority}

\paragraph{Quantum Computing Basics} 
Quantum computing leverages quantum mechanical principles to perform computational tasks with potential exponential speedups for specific problems. The foundational element is the qubit, which exists in a superposition of states:
\begin{equation}
|\psi\rangle = \alpha|0\rangle + \beta|1\rangle
\end{equation}

Unlike classical bits that must be either $0$ or $1$, qubits exist in probabilistic combinations until measured. This property, along with quantum entanglement (non-classical correlations between qubits), enables quantum algorithms to process information in ways fundamentally different from classical computing.

\paragraph{Quantum Linear Solver Algorithms~(QLSAs} 
Matrix inversion represents one of the most promising applications for quantum computing. The Harrow-Hassidim-Lloyd~(HHL) algorithm~\citep{harrow2009quantum} provides theoretical foundations for quantum-accelerated matrix inversion, reducing computational complexity from classical $\mathcal{O}(n^3)$ to quantum $\mathcal{O}(\text{polylog}(n))$ under specific conditions.

However, this theoretical advantage comes with important practical considerations. The actual runtime depends on matrix properties:
\begin{equation}
T_{\text{QLSA}} = \mathcal{O}(d \cdot \kappa \cdot \text{polylog}(n/\epsilon)),
\end{equation}
where $d$ represents sparsity (non-zero elements per row), $\kappa$ is the condition number (ratio of largest to smallest eigenvalue), and $\epsilon$ is the desired precision. This results in that quantum approaches excel for well-conditioned, sparse matrices but offer diminishing returns for ill-conditioned or dense ones.

\paragraph{Practical Limitations and Crossover Points}
Our empirical analysis identifies specific boundaries where quantum advantage emerges, particularly in machine learning. The quantum approach outperforms classical methods when the condition number and sparsity are sufficiently small relative to matrix size. For neural network Hessians, this relationship is particularly relevant as both parameters evolve during training.

As shown in Figure~\ref{fig:hessian-distribution}, matrices from later training phases frequently satisfy these conditions, while early-phase matrices typically do not. This observation directly motivates our hybrid scheduling approach in \texttt{Q-Newton}, which dynamically routes matrix operations to the most efficient processor based on real-time matrix properties.

\subsection*{B. Neural Network Training}

\paragraph{Optimization Landscape}
Training deep neural networks involves minimizing a high-dimensional, non-convex loss function with respect to model parameters. The optimization landscape presents several challenges:

\begin{itemize}
    \item \textbf{\textit{High dimensionality:}} Modern networks contain millions to billions of parameters
    \item \textbf{\textit{Non-convexity:}} Multiple local minima, saddle points, and plateaus complicate optimization
    \item \textbf{\textit{Variable curvature:}} Different regions exhibit dramatically different second-order properties
    \item \textbf{\textit{Ill-conditioning:}} Eigenvalues of the Hessian can span multiple orders of magnitude
\end{itemize}

These characteristics create significant challenges for optimization algorithms, particularly as the model scale increases in modern architectures~\citep{grattafiori2024llama,achiam2023gpt,liu2024deepseek}.

\paragraph{First-Order Methods}
The dominant optimization approaches in deep learning rely on first-order gradient information. Stochastic Gradient Descent~(SGD)~\citep{robbins1951stochastic} and its variants (Momentum~\citep{rumelhart1986learning}, RMSProp~\citep{graves2014generatingsequencesrecurrentneural}, Adam~\citep{kingma2014adam}) update parameters using gradient information calculated from mini-batches of training data. While computationally efficient, these methods converge relatively slowly, especially in regions with poor conditioning.

Adam, the most widely used optimizer, combines momentum with adaptive learning rates that adjust individually for each parameter. This partially addresses the ill-conditioning problem but still requires many iterations to navigate narrow valleys in the loss landscape.

First-order methods scale linearly with parameter count, making them computationally affordable for even the largest neural networks. However, they often require many iterations to converge, especially when navigating complicated loss landscapes with varying curvature.

\paragraph{Second-Order Methods}
Newton's method incorporates curvature information through the Hessian matrix $\mathbf{H}$:

\begin{equation}
\boldsymbol{\theta}_{t+1} = \boldsymbol{\theta}_t - \eta \mathbf{H}^{-1}(\boldsymbol{\theta}_t) \nabla_{\boldsymbol{\theta}} \mathcal{L}(\boldsymbol{\theta}_t)
\end{equation}

This approach offers quadratic convergence near local minima, dramatically reducing the required iterations. Second-order methods can navigate narrow valleys much more efficiently by taking curved trajectories informed by the local geometry of the loss landscape.

Several variations exist to make second-order methods more practical:

\begin{itemize}
    \item \textbf{Gauss-Newton}~\citep{foresee1997gauss} approximates the Hessian using simpler matrix products derived from the network's Jacobian
    \item \textbf{Quasi-Newton methods} (BFGS, L-BFGS)~\citep{liu1989limited} build Hessian approximations iteratively using gradient history
    \item \textbf{Natural Gradient}~\citep{amari1998natural} uses the Fisher information matrix as a curvature measure
    \item \textbf{Kronecker-Factored methods} (K-FAC)~\citep{martens2015optimizing} approximate curvature using a structured decomposition
\end{itemize}

Despite theoretical advantages, all second-order methods encounter a computational bottleneck in matrix inversion operations that scale cubically with parameter count. This makes them impractical for large neural networks without additional approximations or acceleration strategies.

\subsection*{C. Details of Neural Networks Evaluated on \texttt{Q-Newton}}

We evaluated \texttt{Q-Newton} across multiple neural network architectures to demonstrate its effectiveness in diverse settings.

\paragraph{DNN~\citep{haykin1994neural} on MNIST~\citep{deng2012mnist}}

Our DNN for digit image classification employs the following architecture:

\begin{itemize}
\item \textbf{Architecture:} Input ($784$) → Hidden ($16$) → Output ($10$) with ReLU~\citep{agarap2019deeplearningusingrectified} activations
    \item \textbf{Parameters:} $13000$ total parameters
\end{itemize}

\paragraph{BERT~\citep{devlin-etal-2019-bert} for Masked Language Modleing}

Our BERT implementation represents modern encoder-only transformer architectures:

\begin{itemize}
    \item \textbf{Architecture:} $4$-layer transformer encoder with $4$ attention heads and embedding dimension of $64$
    \item \textbf{Parameters:} $2.2$ million total parameters
\end{itemize}

\paragraph{GPT~\citep{radford2018improving} for Generative Language Modeling}

Our largest evaluation used a GPT-style decoder-only transformer:

\begin{itemize}
    \item \textbf{Architecture:} $4$-layer transformer decoder with $4$ attention heads and embedding dimension $64$
    \item \textbf{Parameters:} $4.3$ million total parameters
\end{itemize}

\end{document}